\documentclass[conference]{IEEEtran}

\IEEEoverridecommandlockouts

\usepackage{float}
\usepackage{booktabs}   
\usepackage{array}  
\usepackage{cite}
\usepackage{amsmath,amssymb,amsfonts,amsthm}
\usepackage{acronym}
\usepackage{makecell}  
\usepackage{multirow}
\usepackage{comment}
\usepackage{algorithmic}
\usepackage{graphicx}
\usepackage{textcomp}
\usepackage{booktabs} 
\usepackage{multirow} 
\usepackage[font=small]{caption}
\usepackage[font=small]{subcaption}
\usepackage{makecell}
\usepackage{threeparttable}
\usepackage{color}
\usepackage{enumitem}   

\usepackage{setspace}
\usepackage[mathscr]{euscript}

\graphicspath{{fig/}}

\newcommand{\bfg}[1]{\boldsymbol{#1}}
\newcommand{\bfb}[1]{\boldsymbol{\rm #1}}

\usepackage{float}
\newcommand{\nx}{n}
\newcommand{\nk}{k}
\newcommand{\Dt}{h}
\newcommand{\mm}{i}

\acrodef{avr}[AVR]{automatic voltage regulator}
\acrodef{qe}[QE]{quantization event}
\acrodef{pss}[PSS]{power system stabilizer}
\acrodef{ode}[ODE]{ordinary differential equation}
\acrodef{dae}[DAE]{differential algebraic equation}
\acrodef{lte}[TE]{truncation error}
\acrodef{pf}[PF]{participation factor}
\acrodef{sg}[SG]{synchronous generator}
\acrodef{sssa}[SSSA]{small-signal stability analysis}
\acrodef{tds}[TDS]{time-domain simulation}
\acrodef{tm}[TM]{Trapezoidal Method}
\acrodef{fem}[FEM]{forward Euler method}
\acrodef{bem}[BEM]{backward Euler method}
\acrodef{fdpf}[FDPF]{fast decoupled power flow}
\acrodef{tg}[TG]{turbine governor}
\acrodef{der}[DER]{distributed energy resource}
\acrodef{agc}[AGC]{automatic generation control} 
\acrodef{pll}[PLL]{phase-locked loop} 
\acrodef{pi}[PI]{Proportional-Integral} 
\acrodef{qss}[QSS]{quantized state system} 
\acrodef{qbi}[QBI]{quantization-based integration}
\acrodef{tdi}[TDI]{Time Domain Integration}
\acrodef{qss1}[QSS1]{first-order quantized state system}
\acrodef{qss2}[QSS2]{second-order quantized state system}
\acrodef{qss3}[QSS3]{third-order quantized state system}
\acrodef{rk4}[RK4]{Runge-Kutta 4}
\acrodef{rk2}[RK2]{Runge-Kutta 2}
\acrodef{rk6}[RK6]{Runge-Kutta 6}
\acrodef{bdf2}[BDF2]{ 2-step Backward Differentiation Formula}
\acrodef{i}[I]{Integral}
\acrodef{pi}[PI]{Proportional-Integral}
\acrodef{p}[P]{Proportional}
\acrodef{liqss}[LIQSS]{Linearly Implicit Quantized State System}
\acrodef{ab}[AB]{Adams–Bashforth}
\acrodef{aiits}[AIITS]{All-Island Irish Transmission System}
\acrodef{lte}[LTE]{local truncation error}
\acrodef{nr}[NR]{Newton Raphson}

\begin{document}
\title{Ill-Conditioned Power Flow Analysis Using a
Quantized State-Based Approach \\  
}

\author{\IEEEauthorblockN{Liya Huang, Federico Milano, and Georgios Tzounas}
\IEEEauthorblockA{\textit{School of Electrical and Electronic Engineering} \\
\textit{University College Dublin}\\
Dublin, Ireland\\
liya.huang@ucdconnect.ie, 
federico.milano@ucd.ie,
georgios.tzounas@ucd.ie}
\thanks{This work is supported by the CETPartnership Joint Call 2024 under project NU-ACTIS, co-funded by the European Commission (grant no. 101069750).  The participation of University College Dublin is funded by the Sustainable Energy Authority of Ireland (ID 24/RDD/1390).}
\vspace{-4mm}
}


\maketitle
 
\begin{abstract}
This paper focuses on power flow analysis through the lens of the Newton flow, a continuous-time formulation of Newton's method.  Within this framework, we explore how quantized-state concepts, originally developed as an alternative to time discretization, can be incorporated to govern the evolution of the Newton flow toward the power flow solution. This approach provides a novel perspective on adaptive step-size control and shows how state quantization can enhance robustness in ill-conditioned cases. The performance of the proposed approach is discussed with the ACTIVSg70k synthetic test system.
\end{abstract}

 
\section{Introduction}
\label{sec:intro}

\subsection{Motivation}
\label{sec:moti}

Power flow analysis is a fundamental task in power system studies, and several numerical methods are available for its solution, e.g., see \cite{milano2008continuous,milano2019implicit,huang2021deepopf, de2023modal, iwamoto2007load,van2002general}. 
Among them, methods based on the continuous Newton framework \cite{milano2008continuous,milano2019implicit} have shown promising robustness for ill-conditioned cases.   Within this framework, Newton iterations are interpreted as the numerical integration of a dynamical system whose equilibrium corresponds to the power flow solution.  Building on this interpretation, this paper explores how concepts from quantized-state methods -- originally proposed as an alternative to time discretization -- can be incorporated to enhance performance and robustness for challenging ill-conditioned and poorly initialized cases.

\subsection{Literature Review}
\label{sec:lite}

From a numerical viewpoint, power flow problems can be classified into four categories: well-conditioned, 
ill-conditioned,
bifurcation point, 
or unsolvable.
Standard Newton's method is effective in well-conditioned cases. 
Saddle-node bifurcations, on the other hand, are commonly detected using continuation methods
\cite{li2008continuation}, while 
maximum loadability optimal power flow formulations \cite{bukhsh2013local}
offer an alternative 
approach.  

This paper focuses on ill-conditioned cases, where a solution exists but standard approaches like Newton's method and \ac{fdpf} \cite{van2002general} struggle to converge.  Such cases are commonly addressed using robust Newton variants, e.g., see \cite{overbye2002power}.  
The continuous Newton framework generalizes these approaches by reformulating the power flow equations into a continuous-time dynamical system, the \textit{Newton flow} \cite{hauser2005continuous}.  This allows the solution to be obtained through numerical integration using any explicit or implicit discretization scheme \cite{milano2008continuous,milano2019implicit}.  For implicit schemes, each integration step requires the solution of a nonlinear algebraic system via Newton iterations, resulting in a double (inner/outer) loop structure.  In this context, adaptive time stepping has also been explored as a means of improving efficiency, e.g., by updating the step size based on local truncation error estimates in explicit schemes or on the convergence of the inner loop in implicit ones.

It is natural to ask whether discretization paradigms beyond conventional time-stepping can be incorporated into this framework and provide enhanced performance and robustness.  The main idea of this paper is that such an alternative paradigm can be provided by \ac{qss} methods \cite{kofman2001quantized, kofman2002second}. 
These methods replace time discretization with state quantization, triggering updates only when a state changes by a prescribed amount, the \textit{quantum}.  
Recent work has provided additional insight into \ac{qss} through a \textit{duality} perspective, in which time is allowed to evolve as a state-dependent variable \cite{huang2025duality}.   
In this paper, we explore how the \ac{qss} logic can be embedded into the Newton flow to realize event-driven step-size adaptation for ill-conditioned power flow analysis.

\subsection{Contribution}
\label{sec:con}

The main contribution of this paper is to introduce \ac{qss} concepts into the continuous Newton framework for power flow analysis, enabling state-event-driven step-size adaptation for solving the associated Newton flow.  In addition, we provide a matrix-pencil–based analysis of explicit and implicit discretizations of the Newton flow, offering insight into their local convergence properties. The proposed approach shows potential to achieve improved performance and robustness for ill-conditioned and poorly initialized cases. 

\subsection{Paper Organization}
\label{sec:orga}

The remainder of the paper is organized as follows.  Section~\ref{sec:cnm} reviews the continuous Newton framework and analyzes the local convergence of its explicit and implicit discretizations under varying step sizes.  Section~\ref{sec:qss} introduces the \ac{qss} concept and its use for event-driven step-size adaptation. Section~\ref{sec:case} presents the case study based on the ACTIVSg70k test system.  Finally, conclusions are drawn in Section~\ref{sec:conclusion}.

\section{Continuous Newton Approach for \\ Power Flow Analysis}
\label{sec:cnm}

This section presents an overview of the continuous Newton approach applied to power flow analysis.  We first outline its formulation and describe explicit and implicit variants. Then, we evaluate local convergence around the power flow solution.

\subsection{Formulation}

Power flow analysis refers to solving the nonlinear system:
%
\begin{equation}
\label{eq:algeb}
\bfg{0} = \bfg{g}(\bfg{y}_o)
\end{equation}
where $\bfg{y}_o \in \mathbb{R}^n$ is the column vector of system variables, including voltage magnitudes and angles at PQ buses, reactive power and voltage angles at PV buses, and active/reactive power at the slack bus. The functions $\bfg{g}(\bfg{y}_o) \in \mathbb{R}^n$ represent the corresponding power flow constraints across the network. 

The main idea of continuous Newton's methods is that the solution of \eqref{eq:algeb} coincides with the steady state of the \textit{Newton flow}, i.e., of the continuous system: 
\begin{equation}
\label{eq:cont}
\bfg g_{\bfg{y}} \; 
{\bfg{y}}' 
= - 
\bfg g(\bfg{y})
\end{equation}
where $\bfg{g}_{\bfg{y}}$ is the Jacobian matrix $\bfg{g}_{\bfg{y}} = {\partial \bfg{g} }/ {\partial \bfg{y}}$.  In other words, power flow analysis is translated into the problem of numerically integrating \eqref{eq:cont} until its equilibrium $\bfg y_o$ is reached. Moreover, since only the steady state is of interest, the trajectory of $\bfg y$ itself is irrelevant -- what matters is whether a discrete approximation of \eqref{eq:cont} converges and how quickly.

The simplest among all discretization schemes is the \ac{fem}.  Application to \eqref{eq:cont} gives:
\begin{align}
\label{eq:fem}
\bfg 0 &= 
\bfg{g}_{\bfg{y}_k}
(
\bfg{y}_{k+1} - 
\bfg{y}_{k} 
)
+
\Dt_k 
\;
\bfg{g}(\bfg{y}_k) 
\end{align}
where $\Dt_k$ is the integration step size.  Obtaining $\bfg{y}_{k+1}$ requires factorizing the matrix $\bfg{g}_{\bfg{y_k}}$ at each step.

Observe that \eqref{eq:fem} corresponds to the $k$-th iteration of Newton's method with damping factor $\Dt_k \in (0,1]$.  In particular, setting $\Dt_k = 1$ recovers standard Newton's method, which is effective when $\bfg{g}_{\bfg{y}_k}$ is well-conditioned, whereas $\Dt_k < 1 $ yields \textit{robust} Newton's method, useful to improve convergence in cases where $\bfg{g}_{\bfg{y}_k}$ is ill-conditioned.

A limitation of \eqref{eq:fem} is that its numerical stability margin shrinks as $\bfg{g}_{\bfg{y}_k}$ becomes ill-conditioned.  This forces small step sizes to achieve convergence.  In this context, implicit schemes such as the \ac{bem} become attractive \cite{milano2019implicit}, as they remain stable even for arbitrarily large step sizes and overdamp fast dynamics, offering  improved robustness.  Applied to the Newton flow \eqref{eq:cont}, \ac{bem} reads: 
\begin{align}
\label{eq:bem}
\hspace{-1mm}
\bfg 0 
 &=
\bfg{g}_{\bfg{y}_{k+1}} \!
(\bfg{y}_{k+1} \! - \bfg{y}_{k} )
+ \Dt_k \bfg{g}(\bfg{y}_{k+1}) 
= \bfg \phi(\bfg{y}_{k+1}) 
\end{align}
System \eqref{eq:bem} is solved at each step using Newton iterations:
\begin{align}
\label{eq:inner}
\bfg{y}_{k+1}^{[\mm+1]} = \bfg{y}_{k+1}^{[\mm]} -\bfg \phi_{\bfg y}^{-1}( \bfg{y}_{k+1}^{[\mm]}) 
\bfg \phi( \bfg{y}_{k+1}^{[\mm]} ) \; , \quad  \mm \in \mathbb{N} \, .
\end{align}

Thus, \ac{bem} leads to a double-loop algorithm:  an \textit{inner} loop, where, for a given $k$, \eqref{eq:inner} is solved to compute $\bfg y_{k+1}$ from $\bfg y_k$; and an \textit{outer} loop, where, once $\bfg y_{k+1}$ is obtained, $k$ is advanced and the process is repeated until $|\bfg y_{k+1} - \bfg y_k| < \varepsilon$.  When this condition is satisfied, \eqref{eq:cont} has reached steady state, i.e., $\bfg{y}' = \bfg 0$, and \eqref{eq:algeb} is solved.  Further implementation details of \eqref{eq:inner} are provided in the case study of Section~\ref{sec:case}.

\subsection{Local Convergence Analysis}
\label{sec:sssa}

We discuss the local convergence of \ac{fem} (i.e., robust Newton's method) and \ac{bem} in the neighborhood of the power flow solution. 
To this end, the discrete-time mappings \eqref{eq:fem}, \eqref{eq:bem} are linearized around $\bfg y_o$, and their convergence is assessed using linear stability theory \cite{9695171, tzounas2023unified}.

\textit{FEM}:  Linearizing \eqref{eq:fem} around 
$\bfg{y}_o$ yields:
\begin{align}
\label{eq:fem_li}
\bfg 0 &= 
\bfg{g}_{\bfg{y}_o}
(
\widetilde{\bfg{y}}_{k+1} - 
\widetilde{\bfg{y}}_{k} 
)
+\Dt_k 
\;
\bfg{g}_{\bfg{y}_o}
\widetilde{\bfg{y}}_{k}
\end{align}
where $\widetilde{\bfg{y}}=\bfg{y} - \bfg{y}_o$. Equivalently:
\begin{align}
\label{eq:fem_li2}
\widetilde{\bfg{y}}_{k+1} 
&= 
(1-\Dt_k) 
\widetilde{\bfg{y}}_{k}
\end{align}
where, in exact arithmetic, $\bfg{g}_{\bfg{y}_o}^{-1} \bfg{g}_{\bfg{y}_o} = \bfb I$ and this also holds numerically as long as $\bfg{g}_{\bfg{y}_o}$ is well-conditioned.  The corresponding $z$-domain matrix pencil \cite{dassios2021robust}
$z \bfb I - (1-\Dt_k)\bfb I$ yields a repeated eigenvalue $z=1-\Dt_k$ with multiplicity $n$. The stability condition $|z|<1$ gives the bound $0 < \Dt_k < 2$.  

\textit{BEM}: Linearizing \eqref{eq:bem} around $\bfg{y}_o$ yields:
\begin{align}
\label{eq:bem_li}
\bfg 0 &= 
\bfg{g}_{\bfg{y}_o}
(
\widetilde{\bfg{y}}_{k+1} - 
\widetilde{\bfg{y}}_{k} 
)
+\Dt_k 
\;
\bfg{g}_{\bfg{y}_o}
\widetilde{\bfg{y}}_{k+1}
\end{align}
Equivalently:
\begin{align}
\label{eq:bem_li2}
(1+\Dt_k) \widetilde{\bfg{y}}_{k+1} 
&=  
\widetilde{\bfg{y}}_{k}
\end{align}
For well-conditioned cases, the corresponding $z$-domain matrix pencil $(1+\Dt_k) z \bfb I - \bfb I$ has a repeated eigenvalue $z = {1}/{(1+\Dt_k)}$ with multiplicity $n$.  Since $|z|<1$ holds $\forall \Dt_k > 0$, the method is locally stable around $\bfg{y}_o$ for any positive step.

\begin{figure}[ht!]
    \centering  
    \begin{subfigure}{0.48\columnwidth}
\includegraphics[width=\linewidth]{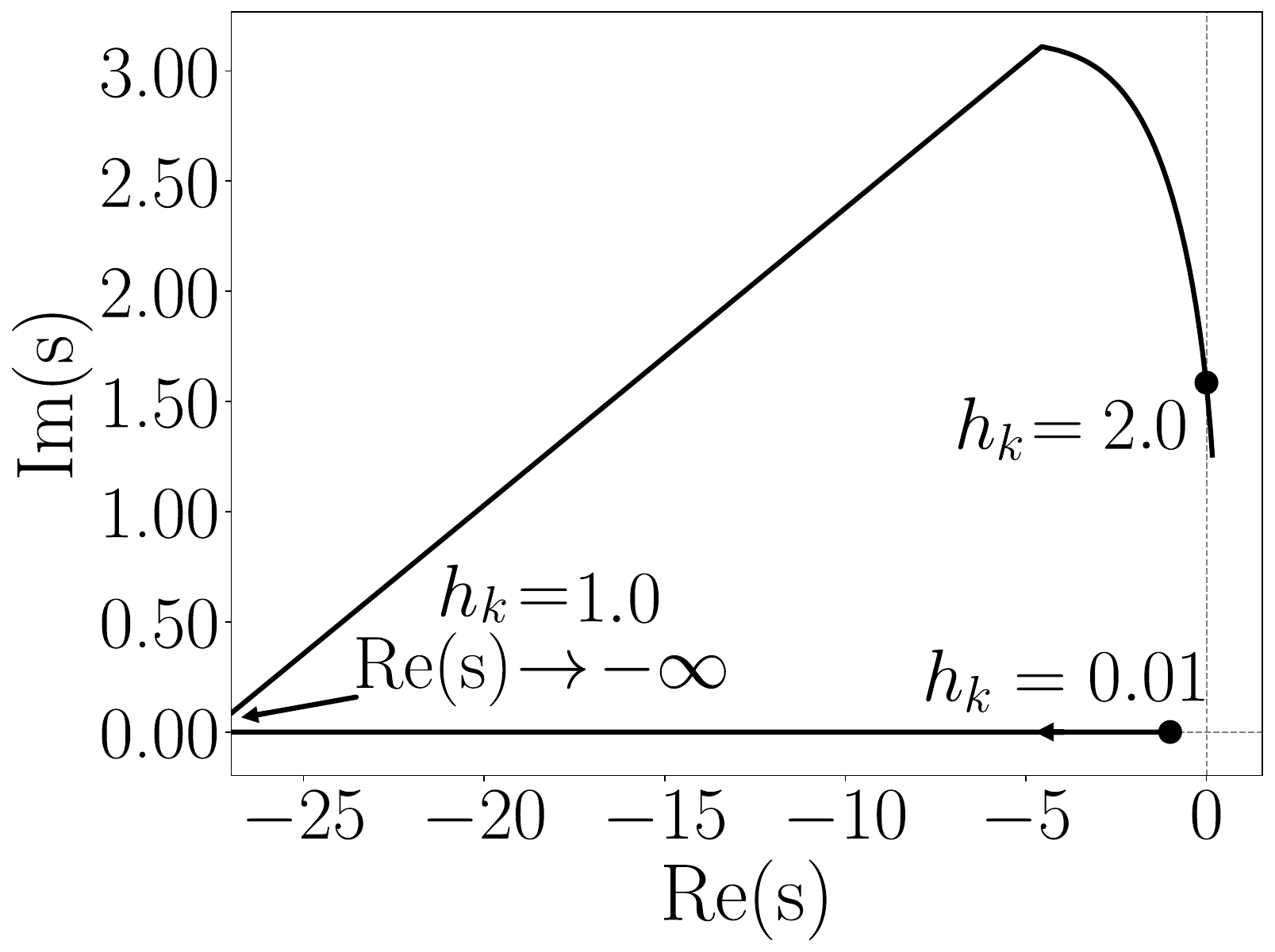}

        \caption{Well-conditioned case.}
        \label{fig:fem_well}
    \end{subfigure}
      \begin{subfigure}{0.48\columnwidth}
   \includegraphics[width=\linewidth]{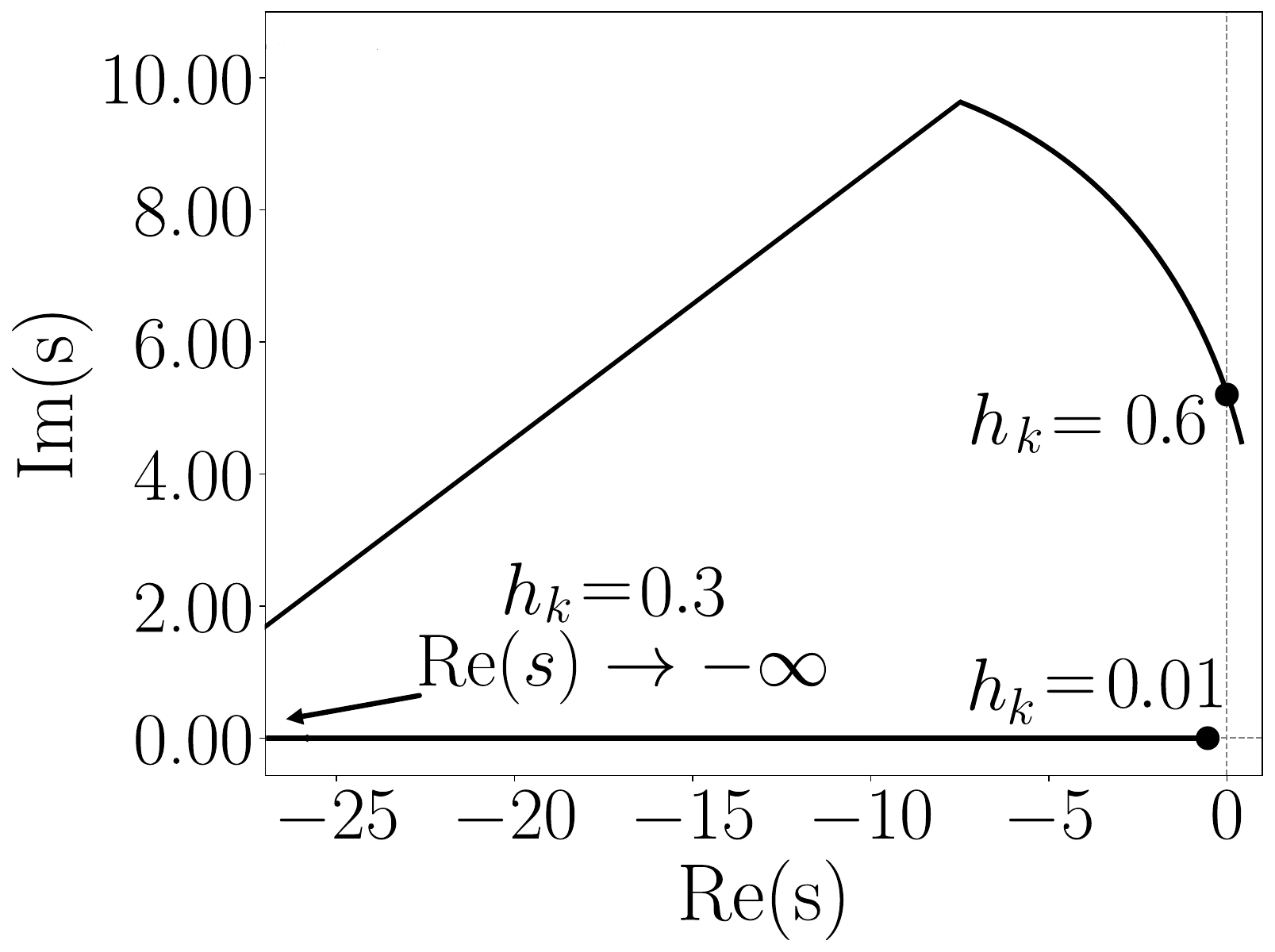}
   
   \caption{Ill-conditioned case.}
        \label{fig:fem_ill}
    \end{subfigure}
    \hfill
    \caption{Local convergence region, FEM.}
    \label{fig:fem:conv}
\end{figure}

\begin{figure}[ht!]
\vspace{-4mm}
    \centering
\includegraphics[width=0.48\columnwidth]{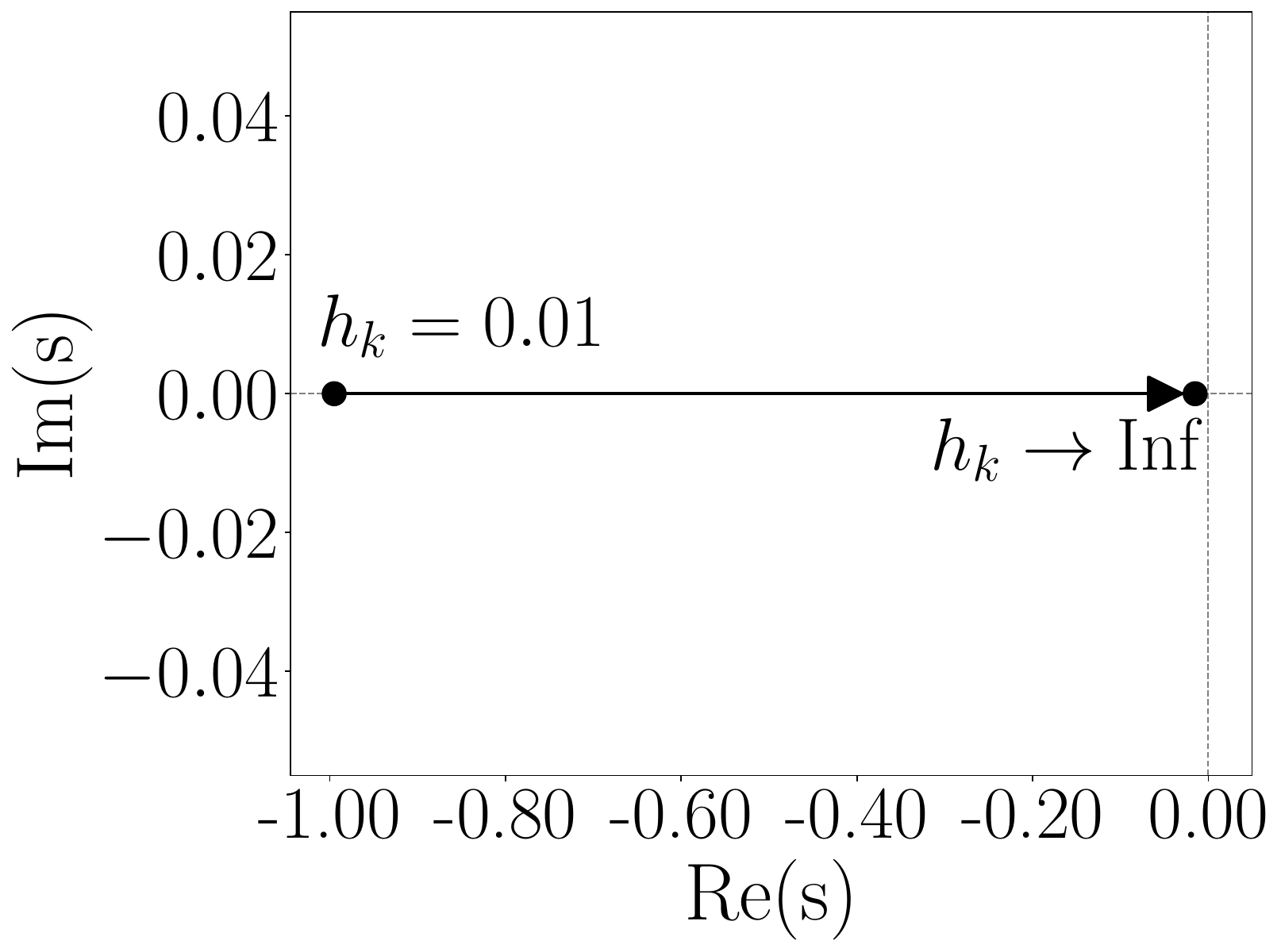}
    \caption{Local convergence region, \ac{bem}.}
    \label{fig:bem:conv}
\end{figure}

Figures~\ref{fig:fem:conv} and  \ref{fig:bem:conv} show the local convergence behavior of FEM and \ac{bem}, respectively, as the step size $\Dt_k$ varies. The results are obtained by first mapping from $z$- to $s$-domain:
 \begin{equation}
   {s}  = \log( {z} )/{\Dt_k} = {\alpha} + j  {\beta}
   \label{eq:mapping}
\end{equation}
Under well-conditioned cases, Fig.~\ref{fig:fem_well} shows that, as expected, convergence of FEM is monotonic for $0<\Dt_k<1$, with the fastest decay rate obtained for $\Dt_k =1$; as $\Dt_k$ increases beyond $1$, behavior becomes oscillatory, and for $\Dt_k>2$, the method diverges.  For \ac{bem}, Fig.~\ref{fig:bem:conv} confirms that convergence is monotonic $\forall\Dt_k>0$, with the decay rate decreasing gradually as $\Dt_k$ increases.  This implies that, near $\bfg y_o$, a larger $\Dt_k$ leads to slower local convergence.

FEM is also known to be sensitive to ill conditioning. 
In particular, when $\bfg{g}_{\bfg{y}}$ is ill-conditioned, its inversion amplifies numerical errors, while residual inaccuracies accumulate across iterations, thus introducing  a non-negligible distortion to the local mapping around ${\bfg{y}_o}$.  To qualitatively show this effect, we consider an illustrative model of the distorted pencil as $z \bfb I - (1-\Dt_k+\epsilon)(1+\eta) \bfb I$, where $\eta$ accounts for imprecision in the factorization of $\bfg g_{\bfg y_o}$; and $\epsilon$ represents the accumulated residual error during Newton iterations.  
For example, with $\eta = 2.33$ and $\epsilon = -0.7$, the local convergence region contracts to $0 < \Dt_k < 0.6$, as shown in Fig.~\ref{fig:fem_ill}.  
In contrast, the implicit nature and strong numerical damping properties of \ac{bem}
greatly reduce the impact of numerical and residual errors, thereby offering enhanced robustness. 
In practice, robustness also depends on the step-size selection.
In the following, \ac{bem} is equipped with a state-event-driven rule for adapting $\Dt_k$.

\section{QSS-Based Step Control}
\label{sec:qss}

In this section, we integrate \ac{qss} concepts into the numerical solution of the power flow problem through the Newton flow formulation \eqref{eq:cont}.  
In the family of \ac{qss} methods, differential equations are solved by quantizing the state variables rather than discretizing time.  The simplest member of this family is the first-order scheme, QSS1~\cite{kofman2001quantized}. Consider the test equation:
\begin{equation}
\label{eq:ode}
y'(t) = f(y(t
)) \ , 
\quad y \in \mathbb{R}
\end{equation}
In QSS1, the derivative $y'(t)$ is evaluated using a piecewise-constant input signal $q(t)$ that approximates $y(t)$:
\begin{equation}
\label{eq:qs}
y'(t) = f(q(t)) 
\end{equation}
where $q(0)=y(0)$.
The signal $q(t)$ updates only when $y(t)$ changes by a fixed \textit{quantum} $\Delta q$:
\begin{equation*}
q(t) = 
\begin{cases} 
y(t)  \ \ \ \ \  \text{if } |y(t)- q(t^-)| \geq \Delta q \\ 
q(t^-) \ \ \; \,  \text{otherwise}
\end{cases}
\end{equation*}
Between two consecutive quantization events, the right-hand side of~\eqref{eq:qs} remains constant, resulting in $y(t)$ being piecewise linear.
Let $t_k$ denote the time of the $k$-th event, where $q(t_k)=y(t_k)$.  
For $t \in [t_k, t_{k+1})$, the state evolves as:
\begin{equation}
 y(t) = y(t_\nk) + (t-t_\nk)  f(y(t_\nk)) \ ,  
 \quad t \in [t_\nk, t_{\nk+1})
\label{eq:xt}
\end{equation} 
The next event occurs at $t=t_{\nk+1}$, when the deviation between $y(t)$ and $y(t_\nk)$ reaches the quantum:
\begin{equation}
|y(t) - y(t_\nk)| = \Delta q  
\label{eq:qe}
\end{equation}
Substituting \eqref{eq:xt} in \eqref{eq:qe} at $t=t_{\nk+1}$,  we get $\Dt_{\nk} | f(y (t_\nk))| = \Delta q $.  Equivalently, the step size $\Dt_{\nk} = t_{\nk+1}- t_\nk$ is:
\begin{equation}
\Dt_{\nk} =  \Delta q /{ |f(y(t_\nk))|}
\label{eq:qss:dt}
\end{equation} 
We now extend the same idea to the Newton flow \eqref{eq:cont}.
The quantized variables are updated according to:
\begin{equation*}
q_j(t) = 
\begin{cases} 
y_j(t)  \ \ \ \ \  \text{if } |y_j(t)-  q_j(t^-)| \geq \Delta q \\ 
 q_j(t^-) \ \ \; \,  \text{otherwise}
\end{cases}
\end{equation*}
where $j$ is the index of vector $\bfg y$, $j \in \{1, \dots, \nx \}$. To compute the derivative of $\bfg{y}(t)$ at the time instant of each quantization event, the quantized signal $\bfg{q}(t)$ is substituted into \eqref{eq:cont}, yielding: 
\begin{equation}
\label{eq:contqss}
\bfg{g}_{\bfg y}(\bfg q(t))\bfg{y}' = - 
\bfg g(\bfg{q}(t))
\end{equation}
$\bfg y'(t)$ is then approximated as:
\begin{equation}
\label{eq:contqss1}
\bfg{y}' =  {\bfg{  f}}(\bfg q(t)) = - [\bfg{g}_{ y}(\bfg q(t))]^{-1}
\bfg g(\bfg{q}(t))
\end{equation}
From this, the step size that determines when each variable reaches its next quantization event can be estimated as:
\begin{equation}
\Dt_{\nk,j} =  \Delta q  / | f_j(\bfg y(t_k))|
\label{eq:hk}
\end{equation}
where $f_j$ is the $j$-th equation of \eqref{eq:contqss1}.  After computing $\Dt_{\nk,j}$ for each equation using \eqref{eq:hk}, we then select the smallest value: 
\begin{equation}
 \Dt_{\nk} = \min\{\Dt_{\nk,j}\} \ , \quad j \in \{1, \dots, \nx \}
 \label{eq:min}
\end{equation}
as the global step size.   
Combination of \eqref{eq:hk} and \eqref{eq:min} produces a QSS-based rule for adaptive step-size selection that integrates naturally with the \ac{bem} discretization of the Newton flow discussed in Section~\ref{sec:cnm}.
Compared with classical adaptive strategies, such as truncation-error controllers or iteration-count heuristics, which typically require tuning multiple parameters, the proposed mechanism is simpler to configure, as it
requires only the choice of $\Delta q$.  Moreover, because the step size is explicitly tied to controlled state variations, the same principle could be leveraged to handle discrete events such as PV–PQ switching.  

\section{Case Study}
\label{sec:case}

This section presents simulation results based on the ACTIVSg70k synthetic test system \cite{birchfield2016grid}.  The system model includes 70,000 buses, 88,207 transmission lines and transformers, and 10,390 generators.  All simulations are carried out in Dome \cite{milano2013python}, on a computer equipped with an Intel Xeon E3-1245 v5 processor, 16~GB of RAM, running a 64-bit Linux~OS.

In the following, we first present base-case results for the original, properly initialized and well-conditioned system.  We then examine performance and robustness under modified, poorly-initialized conditions.  We consider three reference solvers,
namely FEM, FDPF, \ac{rk4} \cite{milano2019implicit}, and four BEM-based variants, including one fixed-step implementation, one with heuristic step adaptation, and two incorporating the proposed QSS-based step control, as summarized in Table~\ref{tab:solvers}.

\begin{table}[ht!]
\centering
\caption{Power flow solver configurations considered.}  
 \label{tab:solvers}
\setlength{\tabcolsep}{3pt}   
\renewcommand{\arraystretch}{0.96}  
\small
\begin{tabular}{@{}lll@{}}  
\toprule  
Notation & Step size & Description  \\
\midrule 
FEM  & Fixed & Robust Newton \\
FDPF & Fixed & Fast decoupled power flow method \\
RK4 & Adap. & Runge Kutta 4 
\\
\midrule
\ac{bem}-J$_1$ & Fixed & \ac{bem} with one inner loop  \\
&&
iteration ($i^{\max}=1$)
\\
\ac{bem}-J & Adap. &
\ac{bem} with with multiple inner 
\\ && 
loop iterations
($i^{\max} > 1$) 
\\ 
\midrule
\ac{bem}-J$_{1}$-QSS & Adap. & 
\ac{bem}-J$_{1}$ with $\Dt_k$ governed by \eqref{eq:min}
\\ 
\ac{bem}-J-QSS & Adap. & 
\ac{bem} with $i^{\max} > 1$ and $\Dt_k$
\\ 
& &
governed
by \eqref{eq:min}
\\
\bottomrule  
\end{tabular}
\end{table}

Unless otherwise stated, the initial step size is set equal to $1$ in all configurations.  For BEM-based solvers, the Hessian term arising in the computation of $\bfg{\phi}_{\bfg y}^{-1}$ in \eqref{eq:inner} is neglected \cite{milano2019implicit}. 
Moreover, for QSS-based schemes, $\Delta q$ is set to 20, while a maximum $h_k^{\max}=8000$ is imposed. These values were chosen after running a large number of tests and were found to provide the best trade-off between accuracy and computational speed.

\subsection{Base-Case Results}

We begin by evaluating the performance of the solver configurations summarized in Table~\ref{tab:solvers} in the base case, corresponding to the original, well-initialized ACTIVSg70k system.
The results are reported in Table~\ref{tab:well-conditioned} and indicate that all methods converge successfully. Among the reference solvers, FEM exhibits the best overall performance. 
Furthermore, the QSS-based methods, \ac{bem}-J-QSS and \ac{bem}-J$_1$-QSS, achieve a significant reduction in the number of iterations compared to their respective counterparts, \ac{bem}-J and \ac{bem}-J$_1$, resulting in a modest speedup. 
Among all tested solvers, \ac{bem}-J$_1$-QSS has the fastest overall convergence. 
For completeness, we note that \ac{rk4} adapts its step based on an estimate of the local truncation error, while \ac{bem}-J adjusts it heuristically according to the convergence rate of the inner Newton loop.

\begin{table}[ht!]
\centering
\caption{Base-case statistics.}  
 \label{tab:well-conditioned}
\setlength{\tabcolsep}{3pt}   
\renewcommand{\arraystretch}{0.96} 
\small
\begin{tabular}{@{}lccc@{}}  
\toprule  
Method & Main loop  & Inner loop & CPU time~[s] \\
\midrule 
FEM  & 25 &  -- & 5.89   \\
FDPF &  12  &  -- &   6.93  \\
 RK4  &  19 &  -- &   \!\!10.24 \\
 \ac{bem}-J$_{1}$ &  25 &  1 & 4.61   \\
\ac{bem}-J  & 24  & 139 & \!\!18.69  \\
\ac{bem}-J$_{1}$-QSS  &  11  & 1&  2.82  \\
\ac{bem}-J-QSS  &  7 &126 &  \!\!17.39 \\
\bottomrule  
\end{tabular}
\end{table}

\begin{figure}[ht!]
    \centering
    \begin{subfigure}{0.49\columnwidth}
        \centering        \includegraphics[width=\linewidth]{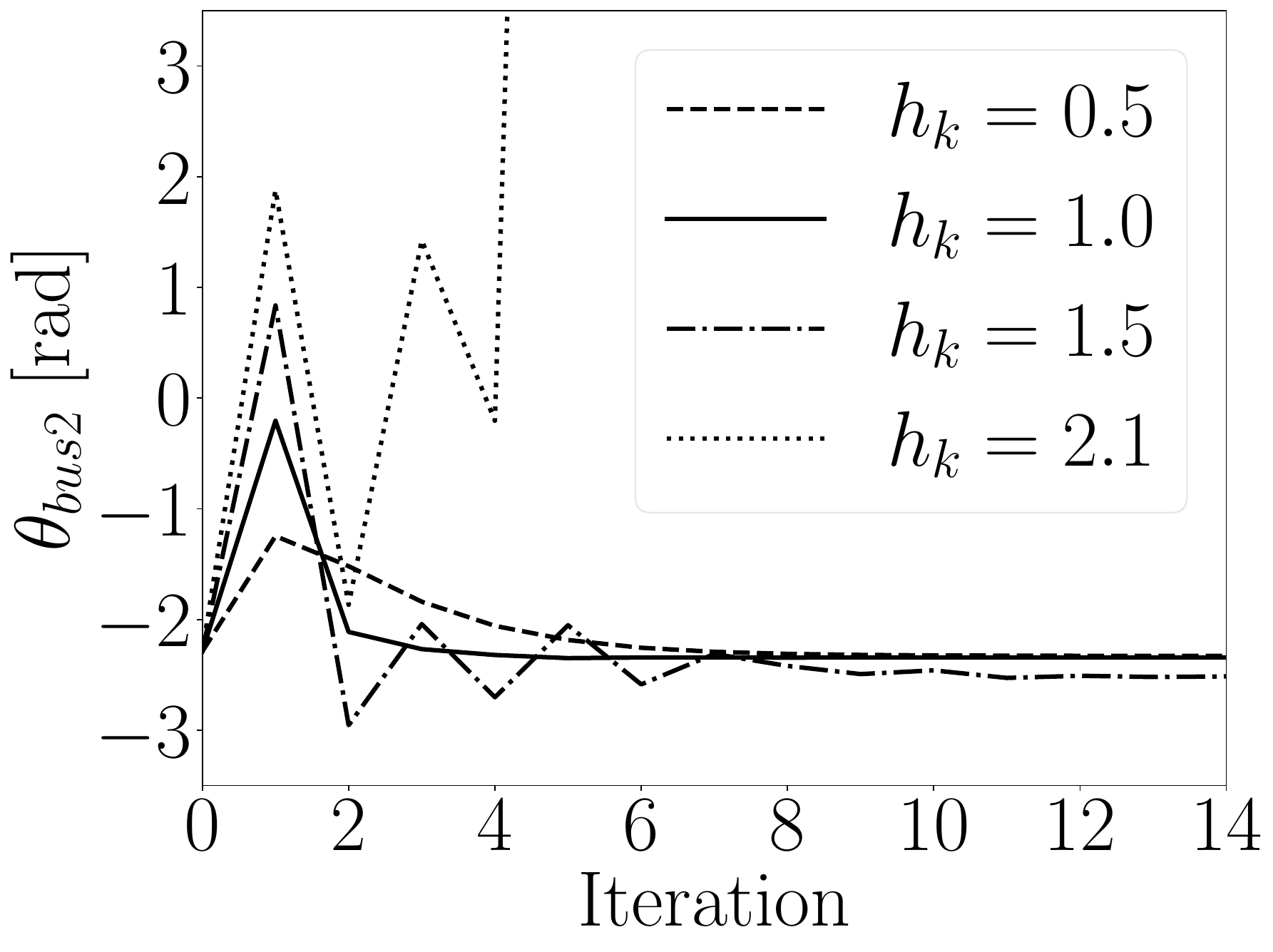}
        \caption{FEM.}
        \label{fig:sub1_robust}
    \end{subfigure}
    \hfill
    \begin{subfigure}{0.49\columnwidth}
    \centering    \includegraphics[width=\linewidth]{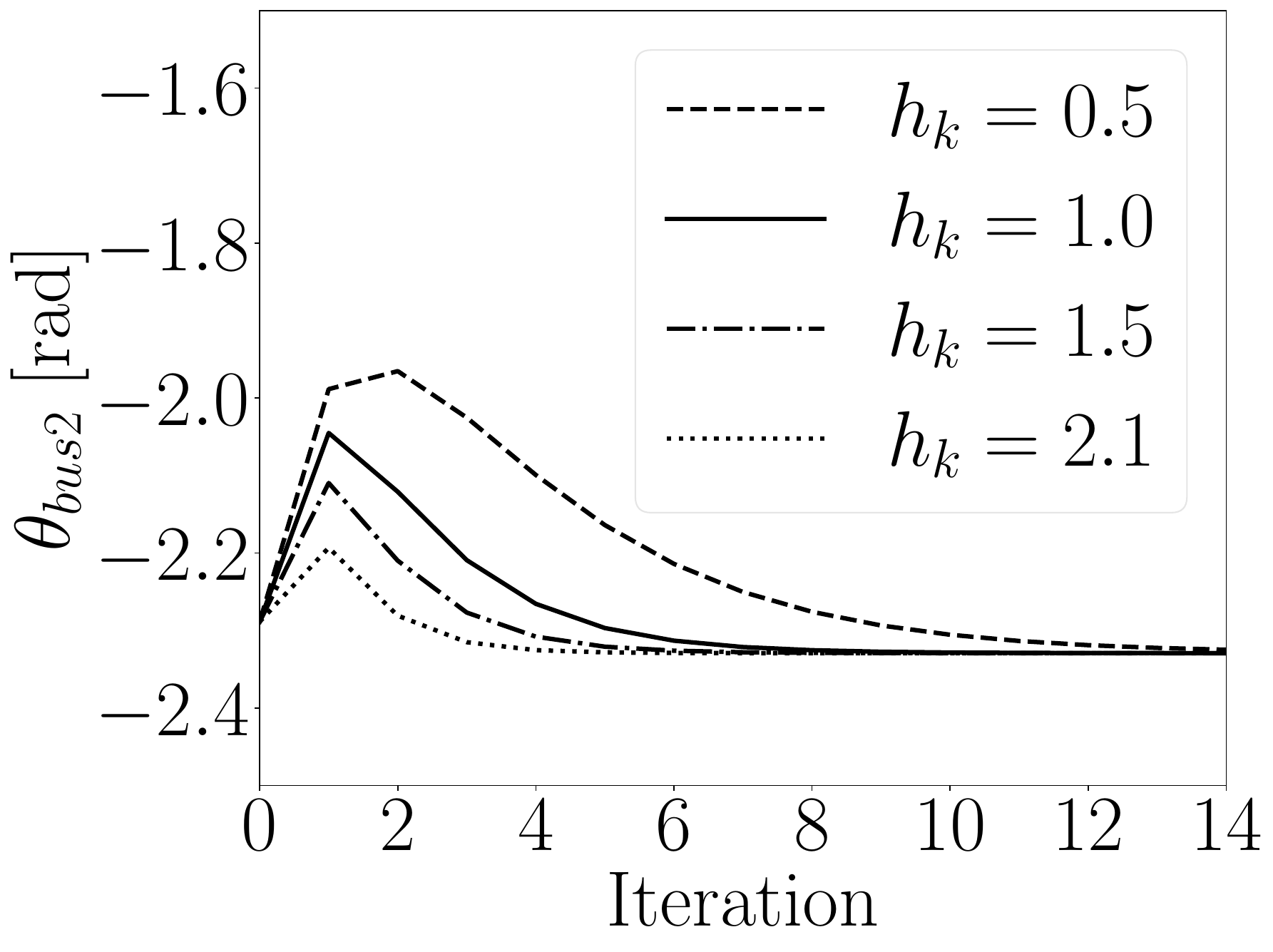}
    \caption{BEM-J with fixed step size.}
    \label{fig:sub2_bemj}
    \end{subfigure}

    \caption{Well-initialized case: FEM and \ac{bem}-J iterations.}
    \label{fig:iteration_two_welll}
\end{figure}

\begin{figure}[ht!]
    \centering
    \begin{subfigure}{0.49\columnwidth}
        \centering        \includegraphics[width=\linewidth]{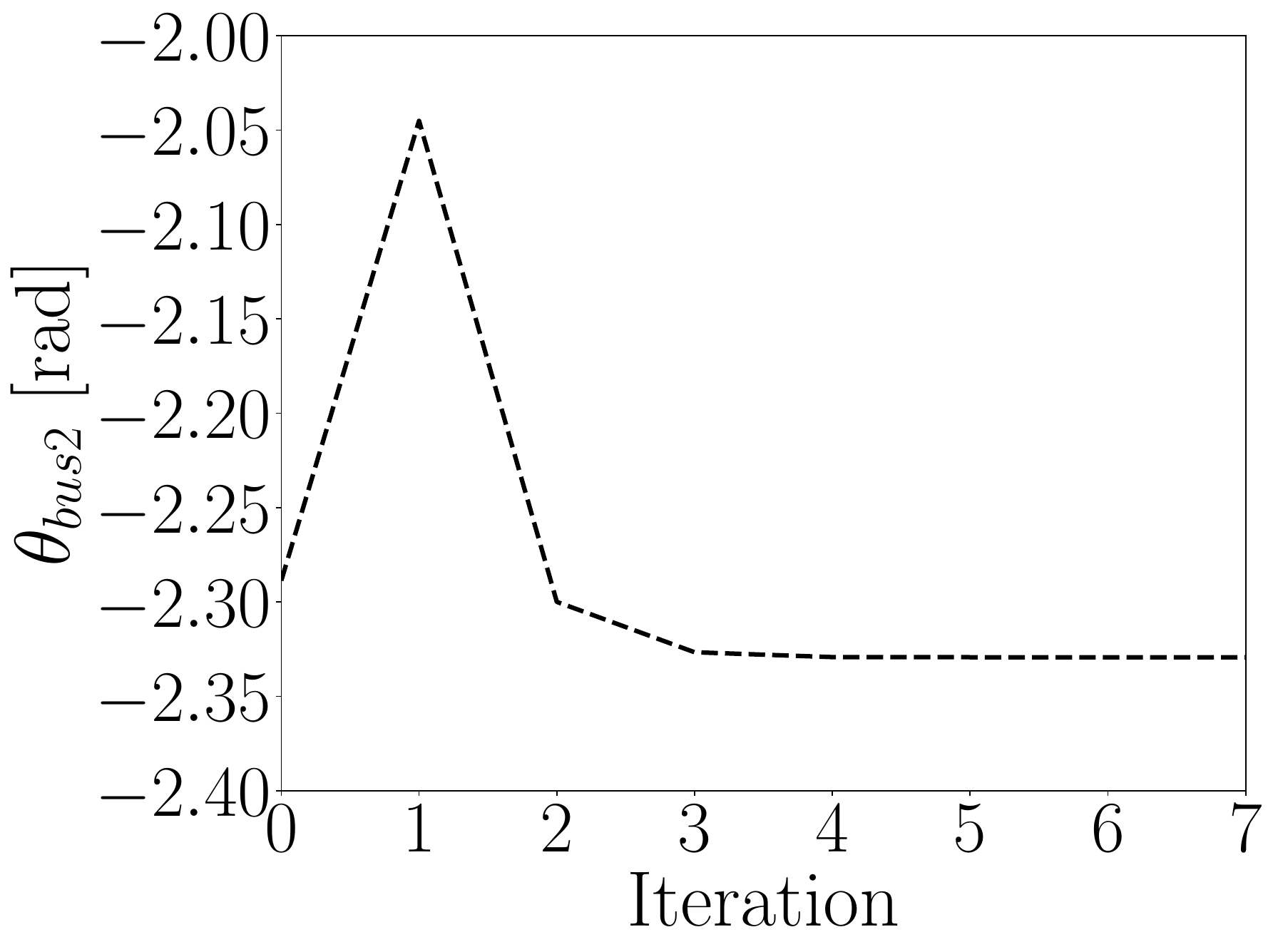}
        \caption{\ac{bem}-J-QSS.}
        \label{fig:bem-j-qss}
    \end{subfigure}
    \hfill
    \begin{subfigure}{0.485\columnwidth}
    \centering    \includegraphics[width=\linewidth]{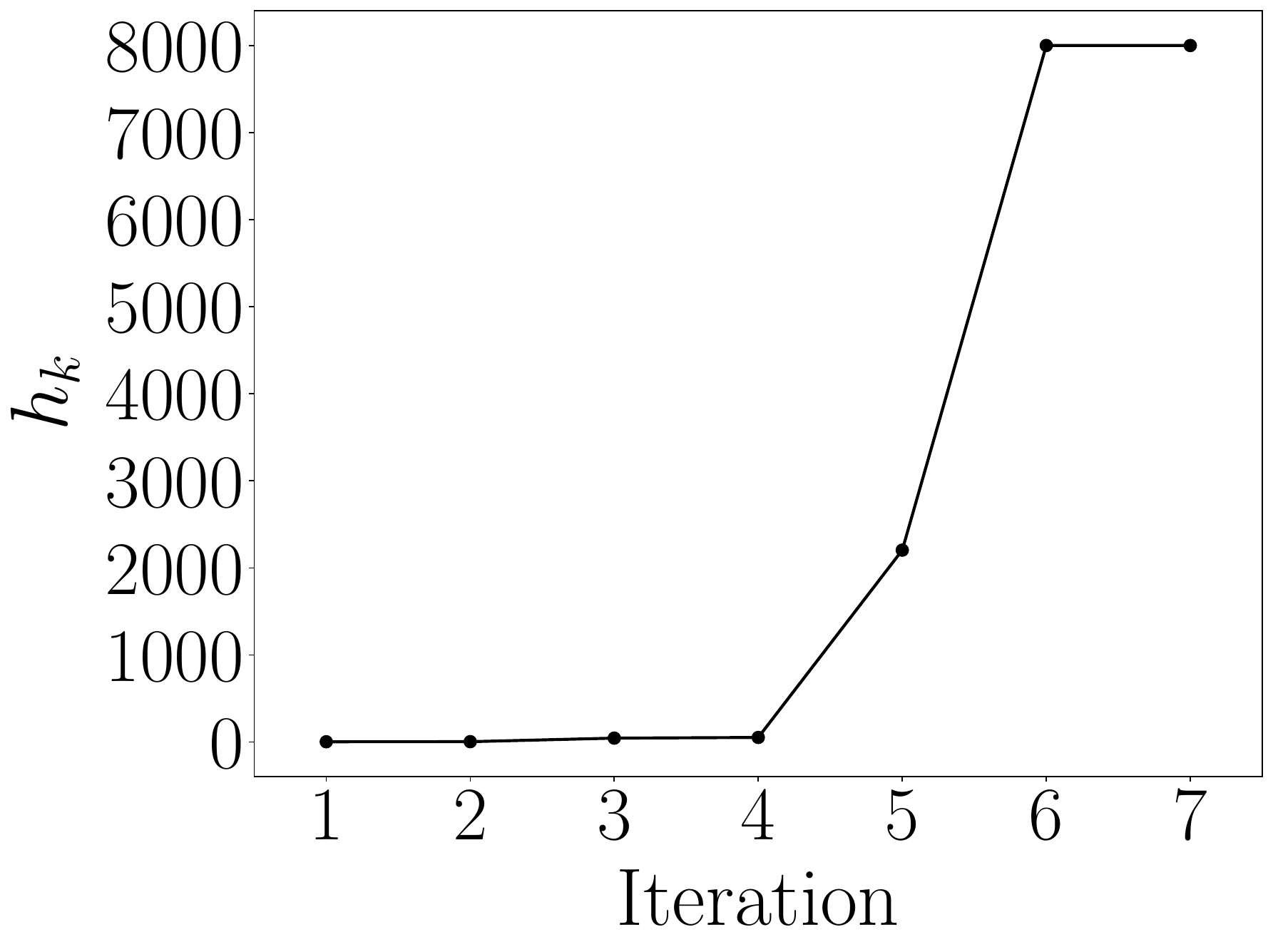}
    \caption{Variation of $\Dt_k$.}
    \label{fig:bem-j-qss-hk}
    \end{subfigure}

    \caption{Well-initialized case: BEM-J-QSS iterations.}
    \label{fig:icnm_qss_well}
\end{figure}

Figure~\ref{fig:iteration_two_welll} shows the iteration trajectory of the voltage phase angle at bus~2 obtained with FEM and \ac{bem}-J.  Both methods are run under fixed steps in this figure to allow a direct comparison.
For FEM, iterations converge fastest at $\Dt_k = 1$, become slower and non-oscillatory for $\Dt_k< 1$, oscillatory for $1 < \Dt_k< 2$, and diverge when $\Dt_k > 2$.  In contrast, \ac{bem}-J converges to the power flow solution without oscillating for all tested step sizes.  These results are consistent with the analysis of Section~\ref{sec:sssa}.
Figure~\ref{fig:icnm_qss_well} shows the trajectory and step-size evolution of \ac{bem}-J-QSS.  It can be seen that \ac{bem}-J-QSS requires fewer iterations than \ac{bem}-J, since $\Dt_k$ increases according to~\eqref{eq:min} as state-derivative magnitudes decrease and the system approaches the steady state.

\vspace{-2mm}

\vspace{-6mm}

\subsection{Performance for Poorly Initialized Conditions}

We examine the performance of the methods in Table~\ref{tab:solvers} under poorly initialized conditions.
Starting from the base case, the initial voltage angles $\bfg{\theta}_{\text{init}}$ are uniformly scaled by a factor $\alpha$ to deteriorate the initialization.

The results obtained for a scaling factor of $\alpha = 1.35$ are summarized in Table~\ref{tab:ill-conditioned}. As shown, this scaling is sufficiently large to cause divergence of the reference solvers.
In contrast, both \ac{bem}-J-QSS and \ac{bem}-J$_1$-QSS converge faster than their respective counterparts, \ac{bem}-J and \ac{bem}-J$_1$.

\begin{table}[ht!]
\centering
\caption{Poorly-initialized-case statistics, $\alpha = 1.35$.} 
\label{tab:ill-conditioned}
\setlength{\tabcolsep}{3pt}   
\renewcommand{\arraystretch}{0.96}  
\small
\begin{tabular}{@{}lccc@{}}  
\toprule  
Method & Main loop  & Inner loop & CPU time~[s] \\
\midrule 
FEM  &  Diverge &  -- & --   \\
FDPF &  Diverge & 
-- & --  \\
RK4 &  Diverge&  -- &  --  \\
\ac{bem}-J$_{1}$   & 25 & 1 &  \!\!4.56 \\
\ac{bem}-J  & 23  & 135 & 
\!\!18.02 \\
\ac{bem}-J$_{1}$-QSS  &12& 1 &  3.67 \\
\ac{bem}-J-QSS  & 7  & 102 & \!\!16.54  \\
\bottomrule  
\end{tabular}
\end{table}

\begin{figure}[ht!]
\centering
\begin{subfigure}{0.495\columnwidth}
\centering
\includegraphics[width=\linewidth]{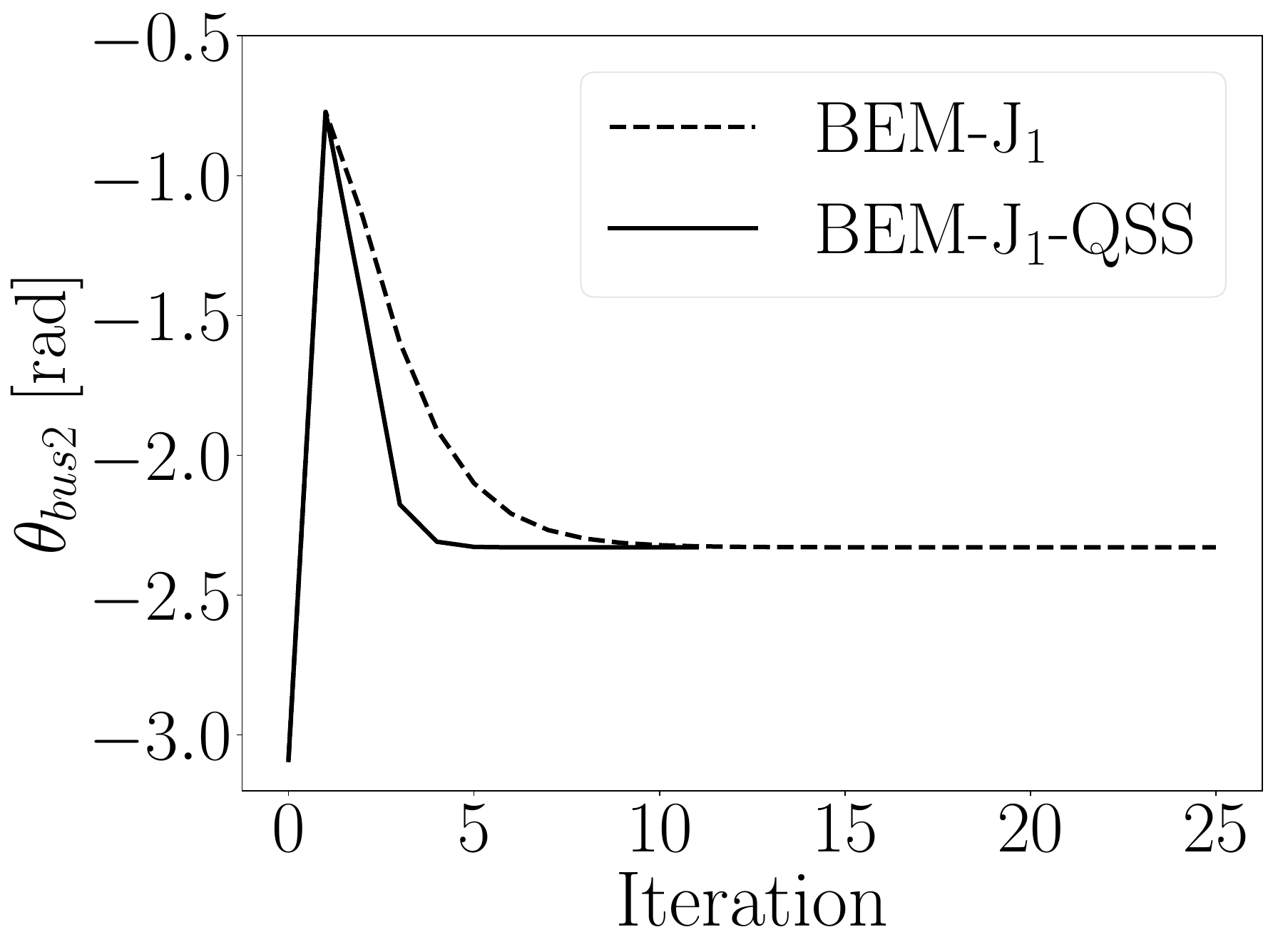}
\caption{Trajectory.}
\label{fig:dt}
\end{subfigure}
\hfill
    \begin{subfigure}{0.49\columnwidth}
        \centering
\includegraphics[width=\linewidth]{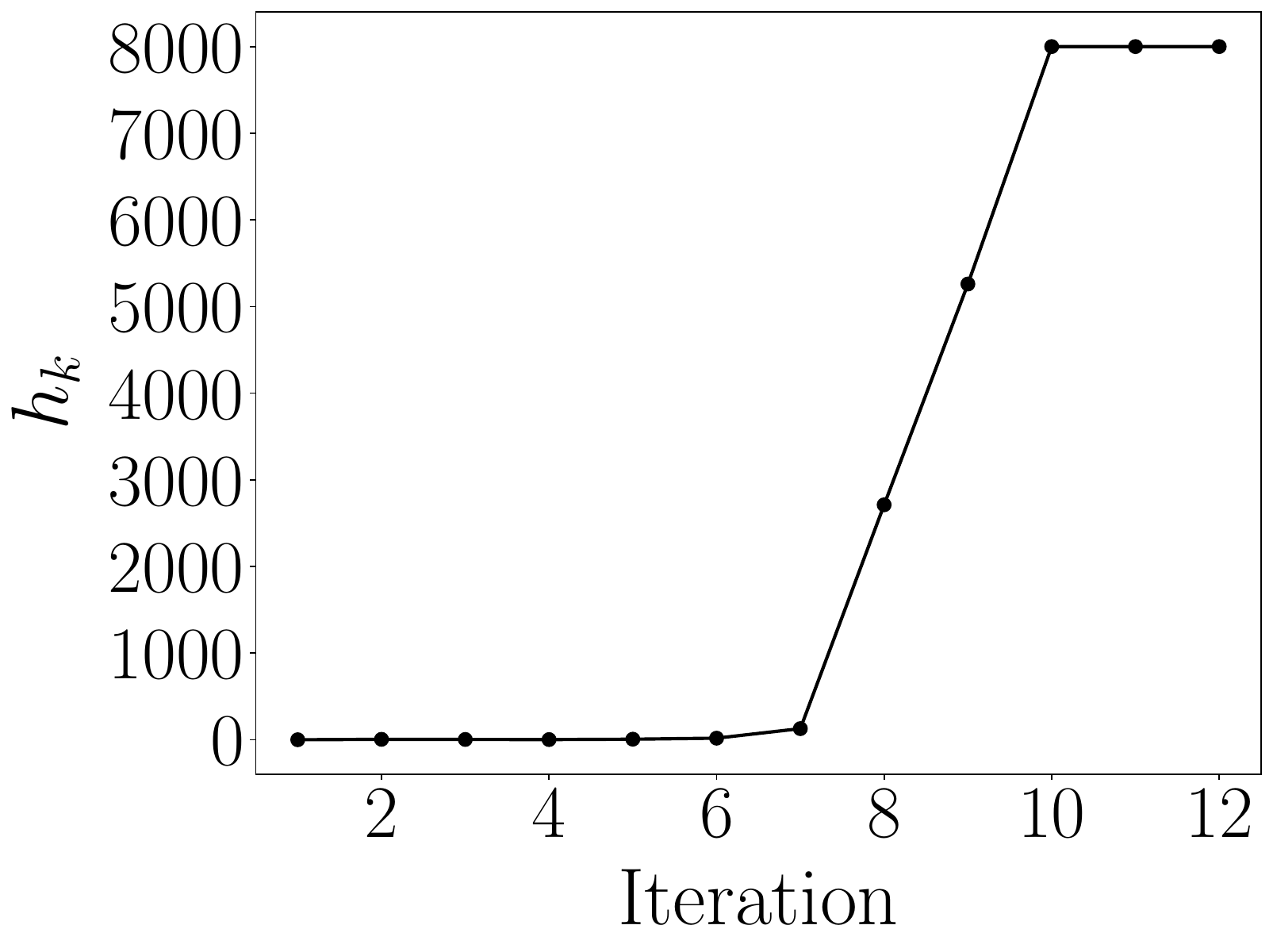}  \caption{Variation of $\Dt_k $, \ac{bem}-J$_{1}$-QSS.}
\label{fig:dt2}
\end{subfigure}
    
\caption{Iterations of $\theta_{bus2}$ with \ac{bem}-J$_{1}$  and \ac{bem}-J$_{1}$-QSS, $\alpha = 1.35$.}
\end{figure}

\begin{figure}[ht!]
\vspace{-2mm}
\centering
\begin{subfigure}{0.48\columnwidth}
\centering
\includegraphics[width=\linewidth]{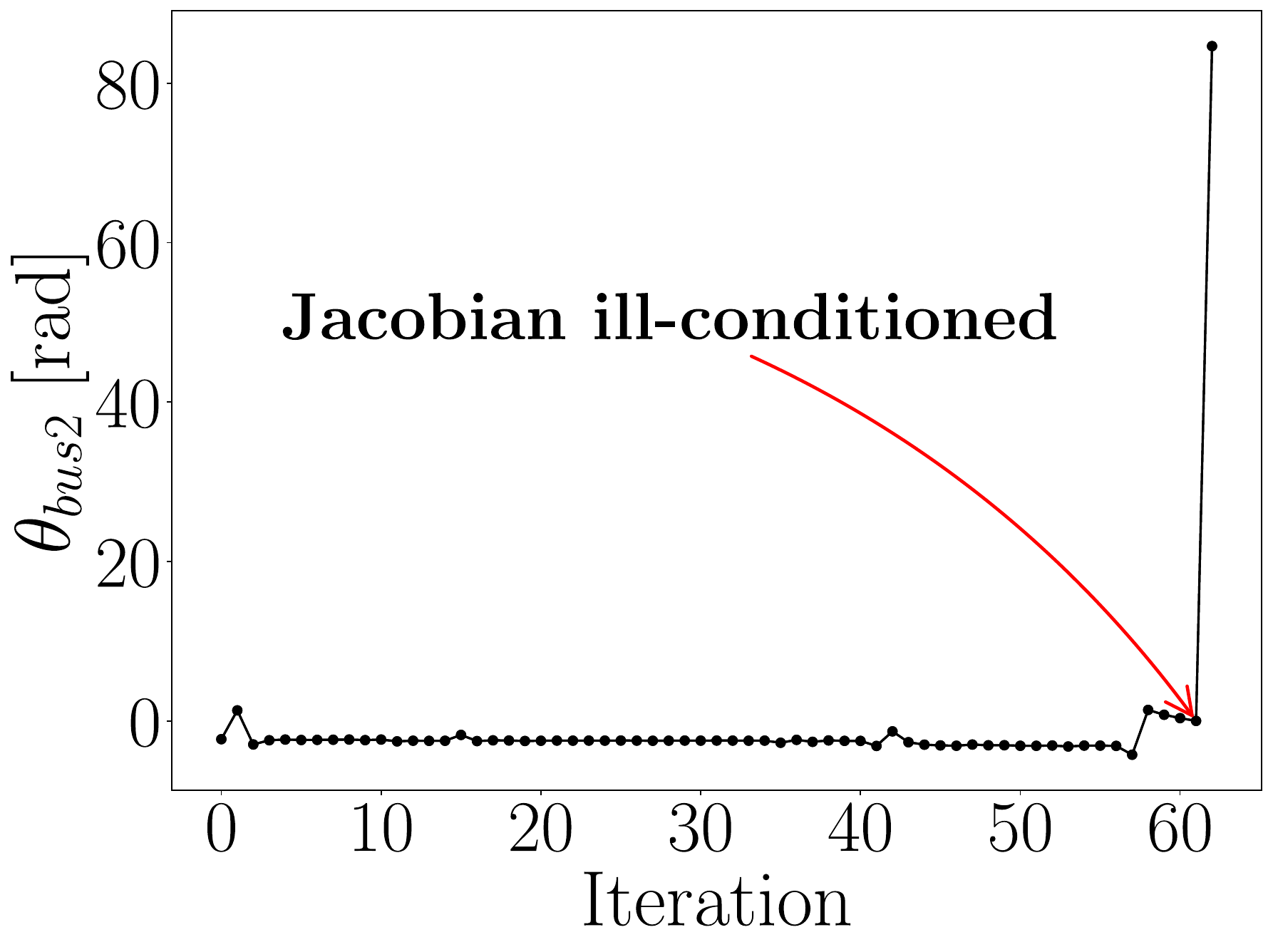}
\caption{\ac{bem}-J$_{1}$.}        \label{fig:dt11}
    \end{subfigure}
    \hfill
    \begin{subfigure}{0.49\columnwidth}
    \centering    \includegraphics[width=\linewidth]{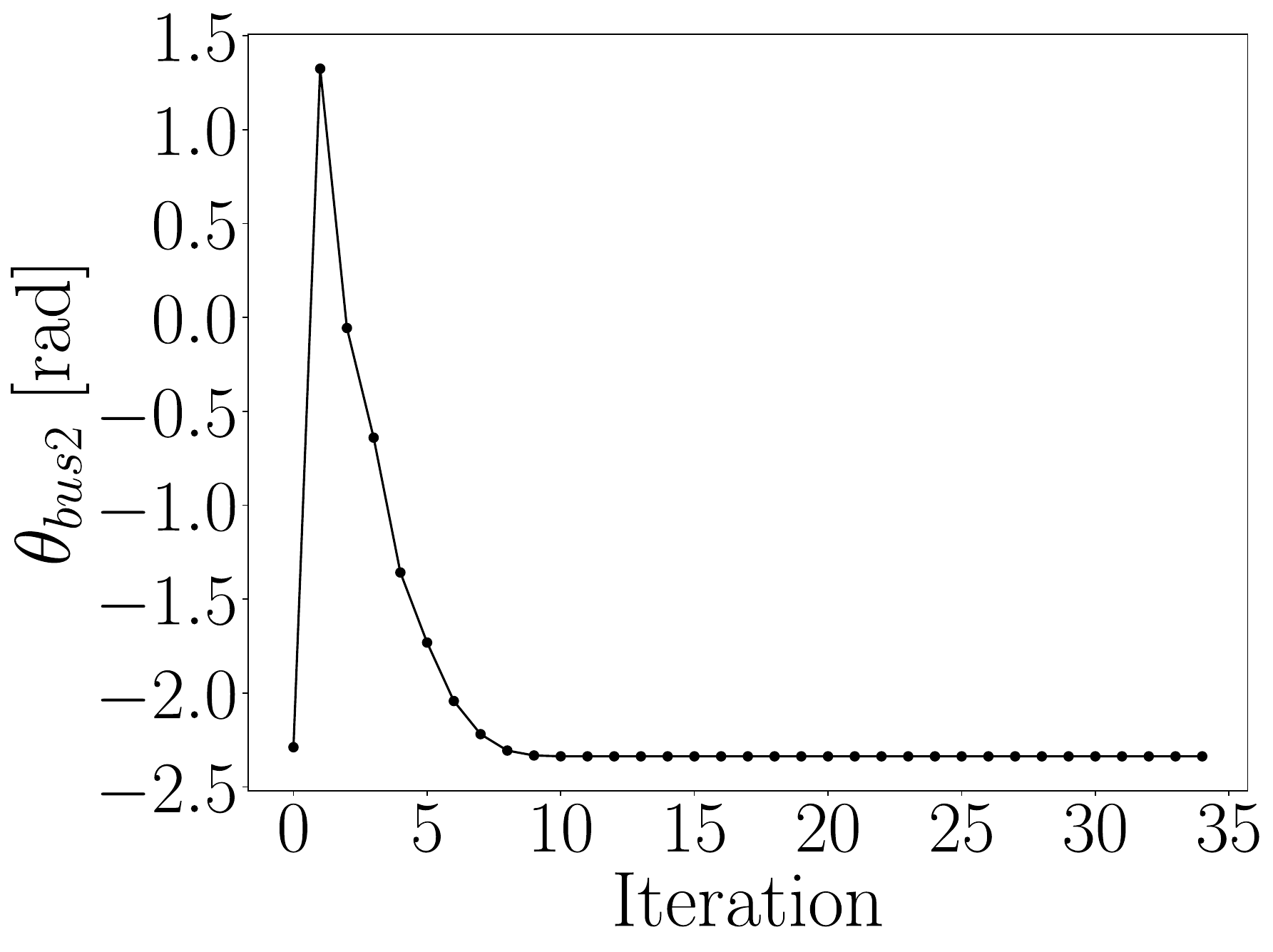}
    \caption{\ac{bem}-J$_{1}$-QSS.}
    \label{fig:dt22}
    \end{subfigure}
\caption{Iterations of $\theta_{bus2}$ with \ac{bem}-J$_{1}$  and \ac{bem}-J$_{1}$-QSS, $\alpha = 1.66$.}
\label{fig:icnm_dt_compare}
\end{figure}

The iteration trajectories of \ac{bem}-J$_1$ and \ac{bem}-J$_1$-QSS for $\alpha = 1.35$ are shown in Figure~\ref{fig:dt}.  Further increasing the scaling factor provides additional insight into their robustness.
In particular, for $\alpha = 1.66$, as shown in Fig.~\ref{fig:dt11}, \ac{bem}-J$_1$ diverges at iteration~62 due to an ill-conditioned Jacobian matrix, whose condition number increases by approximately $10^4$ times compared with its previous value. This issue does not occur in \ac{bem}-J$_1$-QSS.

\begin{figure}[ht!]
    \centering
    \begin{subfigure}{0.48\columnwidth}
        \centering
\includegraphics[width=\linewidth]{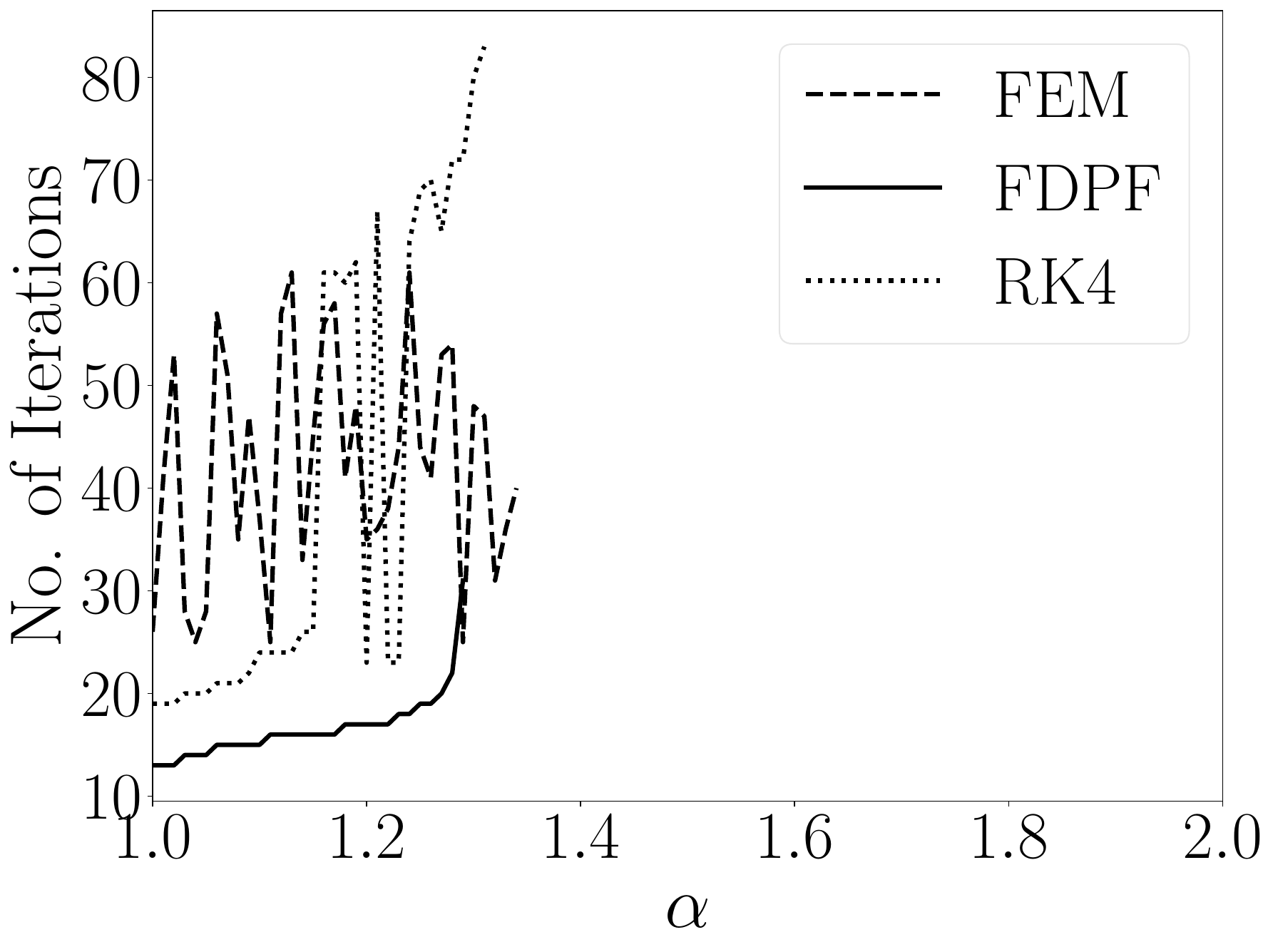}
        \caption{Reference solvers.}
        \label{fig:refs}
    \end{subfigure}
    \hfill
    \begin{subfigure}{0.48\columnwidth}
    \centering    \includegraphics[width=\linewidth]{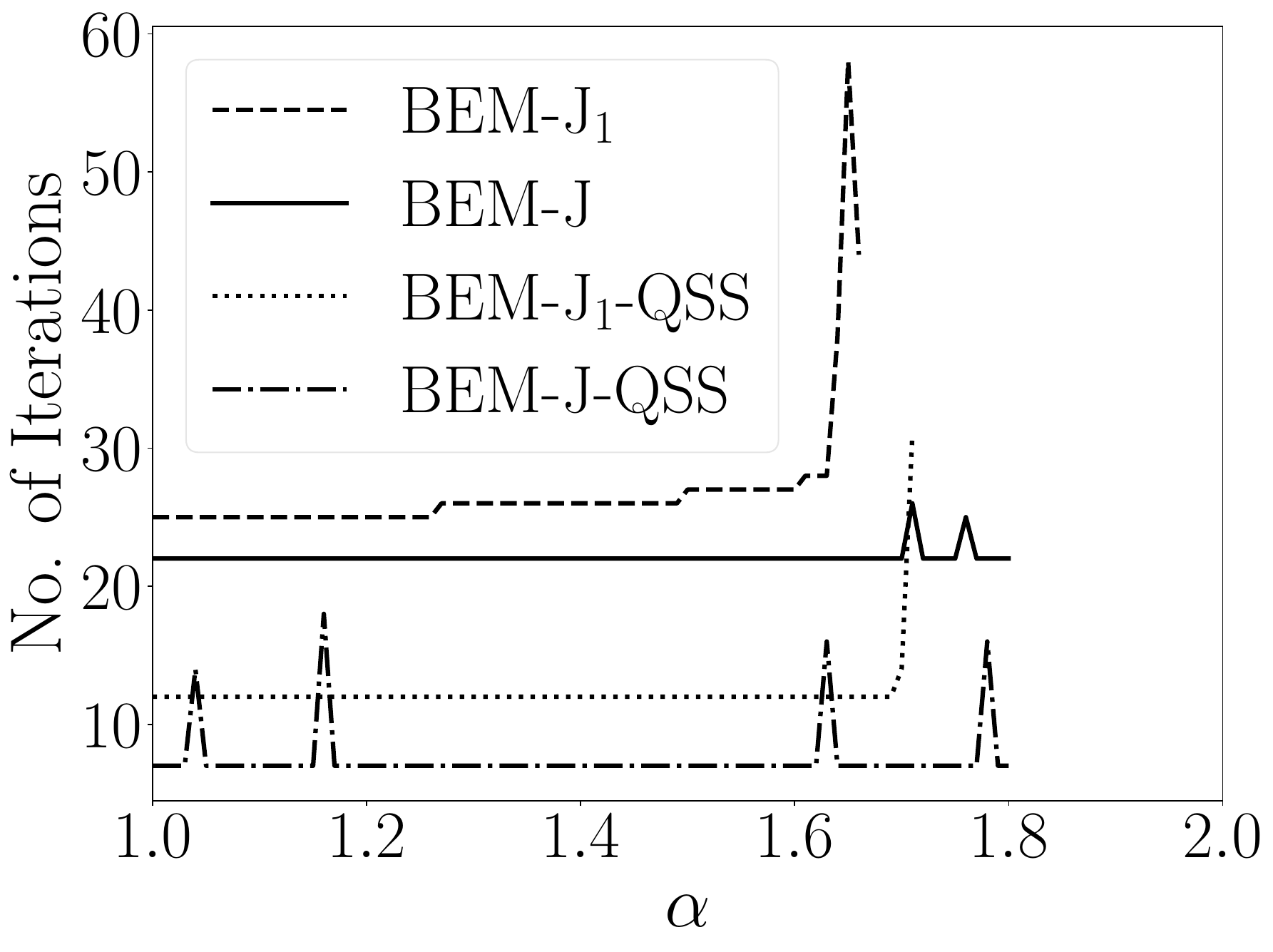}
    \caption{\ac{bem}-based solvers.}
    \label{fig:bems}
    \end{subfigure}
    \caption{Iterations of $\theta_{bus2}$ for all methods, $\alpha \in (1,2]$.}
    \label{fig:robustness}
\end{figure}

To further assess robustness, each method is tested on a set of 500 cases obtained for scaling factors uniformly distributed in the range $\alpha \in (1,2]$.  The percentage of converging cases for all methods in Table~\ref{tab:solvers} is reported in Table~\ref{tab:probability}. As shown, the reference methods (FEM, FDPF, and RK4) exhibit low overall convergence rates under poorly initialized conditions.  \ac{bem}-J and \ac{bem}-J-QSS achieve the highest convergence rates, while \ac{bem}-J$_1$ and \ac{bem}-J$_1$-QSS show reduced robustness due to their inner loop approximation.  The number of iterations required by each method as $\alpha$ varies within $(1,2]$ is shown in Fig.~\ref{fig:robustness}. Each curve terminates at the largest value of $\alpha$ for which the method maintains convergence. 

\begin{table}[ht!]
\centering
\caption{Converging cases under scaled initial voltage angles.}  
\label{tab:probability}
\setlength{\tabcolsep}{3pt}   
\renewcommand{\arraystretch}{0.96}  
\small
\begin{tabular}{@{}lc|lc@{}}  
\toprule  
Method & Convergence [\%] 
& Method & Convergence [\%] \\
\midrule 
FEM & 39.80 & \ac{bem}-J & 82.25
\\
{FDPF} & {32.78} & \ac{bem}-J$_1$-QSS &  68.32
\\
{RK4} &  {33.40}  & \ac{bem}-J-QSS & 82.28
\\
\ac{bem}-J$_1$ & 65.93
\\
\bottomrule  
\end{tabular}
\end{table}

\section{Conclusion}
\label{sec:conclusion}

This paper integrates \ac{qss} concepts into the continuous Newton framework for power flow analysis.  By leveraging QSS-based state-update logic, the proposed approach enables event-driven adaptive step-size control, improving both convergence speed and robustness in ill-conditioned cases.  Simulation results show that it accelerates convergence relative to fixed-step and heuristically tuned \ac{bem} variants and enhances robustness when combined with \ac{bem}-J$_1$.  Future work will examine the use of QSS-based event handling within the same framework to manage switching events during iterations triggered by operational limit violations.
 
\bibliographystyle{IEEEtran}
\bibliography{references}

\end{document}